# The Future of Decoding Non-Standard Nucleotides: Leveraging Nanopore Sequencing for Expanded Genetic Codes

Hyunjin Shim[1, a)]

*Author Affiliations*
[1]*Department of Biology, California State University, Fresno, 5241 N Maple Ave, Fresno, CA 93740, USA*

*Author Emails*
[a)] Corresponding author: jinenstar@gmail.com

**Abstract.** Expanding genetic codes from natural standard nucleotides to artificial non-standard nucleotides marks a significant advancement in synthetic biology, with profound implications for biotechnology and medicine. Decoding the biological information encoded in these non-standard nucleotides presents new challenges, as traditional sequencing technologies are unable to recognize or interpret novel base pairings. In this perspective, we explore the potential of nanopore sequencing, which is uniquely suited to decipher both standard and non-standard nucleotides by directly measuring the biophysical properties of nucleic acids. Nanopore technology offers real-time, long-read sequencing without the need for amplification or synthesis, making it particularly advantageous for expanded genetic systems like Artificially Expanded Genetic Information Systems (AEGIS). We discuss how the adaptability of nanopore sequencing and advancements in data processing can unlock the potential of these synthetic genomes and open new frontiers in understanding and utilizing expanded genetic codes.

## INTRODUCTION

To understand the encoding of biological information, the beginning of life on Earth is a topic that delves into the origins of biological molecules and their assembly into living organisms. One pivotal experiment that tested the theory of primordial soup, which suggests that life began in a "soup" of organic compounds in early Earth's oceans, was conducted by Miller and Urey in 1952 [1]. The Miller-Urey experiment simulated the conditions thought to be present on early Earth by combining water, methane, ammonia, and hydrogen in a closed system and exposing the mixture to electrical sparks to mimic lightning. After a week, they found that several organic molecules, including amino acids, the building blocks of proteins, had formed. This groundbreaking experiment provided the first experimental evidence that simple organic molecules necessary for life could be synthesized from inorganic precursors under prebiotic conditions, supporting the idea that life could have arisen from non-living chemical substances through natural processes [2].

Biological information is encoded in a sequence of biochemical molecules called nucleic acids, which include DNA (deoxyribonucleic acid) and RNA (ribonucleic acid). These nucleic acids are composed of long chains of nucleotides, each consisting of a sugar molecule, a phosphate group, and a nitrogenous base [3]. In DNA, the four nitrogenous bases pair specifically (A with T, and C with G) to form the double helix structure, while RNA contains uracil (U) instead of thymine (T). The sequence of these bases along a nucleic acid strand constitutes the genetic code, which is read in sets of three bases known as codons [4]. Each codon specifies a particular amino acid or signals a start or stop in protein synthesis. Thus, the linear arrangement of nucleotides in nucleic acids dictates the synthesis of proteins, which in turn determine the structure and function of cells, ultimately encoding the genetic instructions for the development, growth, function, and reproduction of all living organisms [5].

For decoding this biological information, the traditional genetic code, consisting of four standard nucleotides (A, T, G, C), has long defined the boundaries of genomic research. However, with the advent of Artificially Expanded Genetic Information Systems (AEGIS) and other non-standard nucleotides, the genetic code is being expanded beyond these four standard nucleotides [6]. Non-standard nucleotides offer the potential to increase the information density and functionality, enabling novel applications in biotechnology, synthetic biology, and even considerations in extraterrestrial exploration [7]. Yet, deciphering these xeno nucleic acids (XNA) sequences requires cutting-edge technologies capable of encoding, sequencing, and decoding these biochemical variations (Figure 1).

Nanopore sequencing, given its ability to directly read native nuclei acids without amplification or synthesis [8,9], offers an unparalleled tool for interpreting these extended genetic codes. Nanopore sequencing represents a groundbreaking innovation in long-read sequencing technology, inspired by the natural mechanisms of bacterial pores [10]. Unlike short-read technologies based on sequencing by synthesis (SBS), this long-read technology utilizes protein nanopores embedded in a synthetic membrane [11,12]. These nanopore sensors measure disruptions in ionic current as nucleic acids pass through nanopores, and these signals are processed using machine learning algorithms to decode biological information to human-readable signals [13]. In this perspective, we discuss nanopore sequencing as a uniquely suited technology to read both standard and non-standard nucleotides, as it directly captures the physical and chemical properties of each molecule.

## MAIN

### Decoding Biochemical Molecules into Human-Readable Signals

The theoretical foundation enabling nanopore sequencing of nucleic acids rests on the principles of biochemistry, biophysics, and engineering [9]. At the heart of the technology is the nanopore, a nanoscale hole typically formed by transmembrane proteins such as alpha-hemolysin or other nanopores embedded within an electrically resistant membrane [12]. When an electric potential is applied across the membrane, ions flow through the nanopore, creating a measurable ionic current. As a single-stranded DNA or RNA molecule passes through the nanopore, it causes a transient disruption in the ionic current, with each nucleotide producing a distinct change in the current amplitude. These current changes, captured in real-time, form a raw signal that directly corresponds to the nucleotide sequence of the passing molecule. Recent progress in single-molecule protein sequencing using nanopores demonstrates the ability of nanopore sequencing to read biological information encoded in a different type of biochemical molecules [13–16].

The second critical component is the computational interpretation of these signals, known as basecalling. The raw ionic current signals are complex and require sophisticated algorithms to translate them into nucleotide sequences [17,18]. Advanced computational methods, including machine learning models like neural networks, are trained to recognize and decode the patterns in the current disruptions caused by different nucleotide sequences [17]. These algorithms account for various factors such as the speed of translocation and the ionic environment, ensuring accurate basecalling even with the inherent noise in the signal. Additionally, the technology leverages adaptive sampling, where the sequencing process can be dynamically adjusted based on real-time analysis, allowing for selective sequencing of regions of interest [19,20]. This intricate interplay between precise biophysical measurement and advanced computational processing enables nanopore sequencing to provide long, continuous reads and detailed insights into biological information, including epigenetic modifications.

The synergistic impact of artificial neural networks and parallel computing is revolutionizing the processing of high-throughput genetic data generated by nanopore sequencing [21]. Artificial neural networks, particularly through deep learning-based basecalling models, significantly enhance the accuracy and speed of translating raw electrical signal data into nucleotide sequences [17]. These models leverage vast amounts of training data to improve prediction accuracy, handling the complex signal patterns produced by nanopore technology. Parallel computing further amplifies this capability by distributing computational tasks across multiple processors, dramatically reducing processing times and enabling real-time data analysis [22]. This combination facilitates adaptive sequencing, where the sequencing process can dynamically adjust based on real-time data analysis, optimizing read output by targeting specific regions of interest. Together, these advancements not only streamline the basecalling process but also enhance the overall efficiency and effectiveness of genomic studies, paving the way for breakthroughs in genomics, personalized medicine, and diagnostics [10,11].

Additional bioinformatics analysis of nanopore reads involves several key steps beyond initial basecalling and classification, aimed at extracting deeper biological insights [23]. One critical step is variant calling, which involves identifying genetic variations such as single nucleotide polymorphisms (SNPs), insertions, deletions, and structural variations within the sequenced genomes. Variant calling from nanopore data allows researchers to track the emergence and spread of mutations, identify drug-resistance genes, and understand genetic diversity within populations [24,25]. Comparative genomics can be employed to identify unique genetic features and evolutionary adaptations of these organisms by comparing their genomes across different environments and conditions.

Pangenome analysis can further elucidate the core and accessory genes, shedding light on the functional diversity and evolutionary dynamics within microbial populations [26]. Additionally, integrating multiomics approaches, such as transcriptomics and proteomics, can provide a holistic understanding of gene expression and regulatory mechanisms. Leveraging advanced bioinformatics tools and machine learning algorithms can enhance the accuracy of genome assembly and annotation, facilitating the discovery of novel genes and pathways [9,27,28].

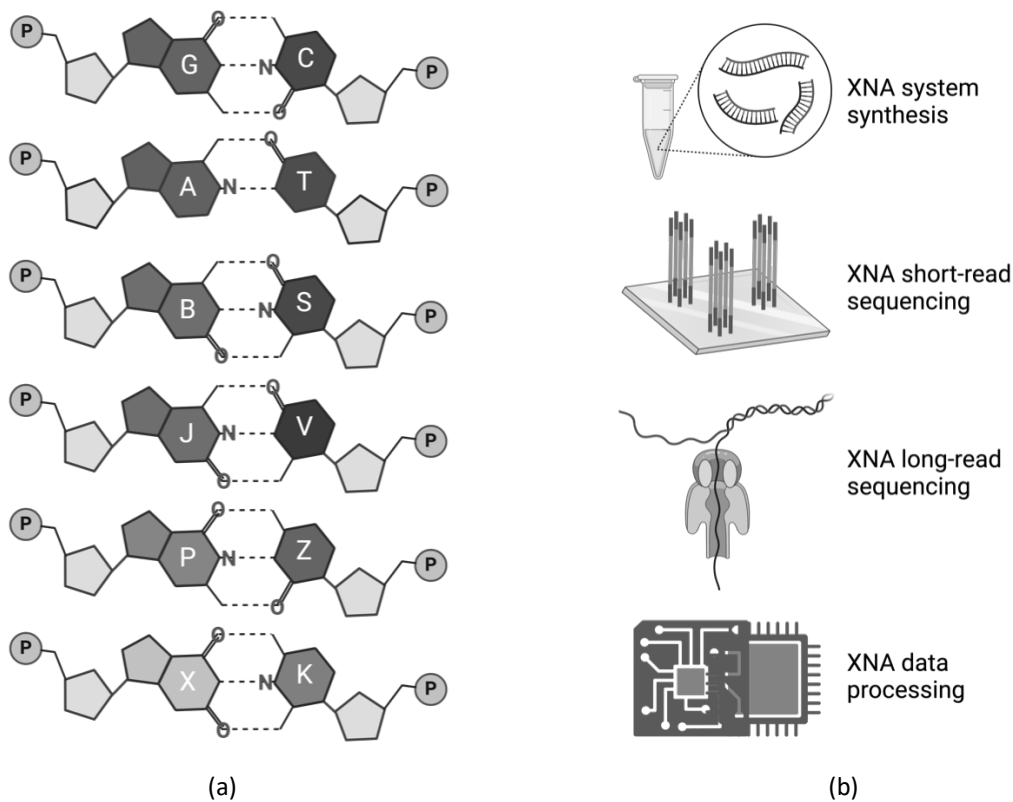

**FIGURE 1.** (a) Expanded genetic code with Artificially Expanded Genetic Information Systems (AEGIS) with hydrogen bond acceptors (O or N) indicated. (b) Future efforts required in multiple disciplines to further explore the potential of expanded genetic codes. Advances in synthetic biology, sequencing, and data processing are necessary to fully apply this synthetic system as genetic toolkits in medicine and biotechnology.

## From AEGIS to Unexplored Frontiers in Synthetic Biology

Since the 1990s, various research groups sought to enhance the functionality of nucleic acids by adding functional groups to the four natural nucleotide bases [29,30]. However, this approach was unable to fully address the limitations of DNA and RNA as scaffolds for generating highly functional binding or catalytic molecules [31]. A recent development involves creating entirely new nucleotide building blocks and incorporating them into the existing XNA pool [32–34]. However, the in vitro selection process of these early systems is still limited by the inability to fully randomize the synthetic extra nucleotides in the initial library, indicating that further refinement is needed for broader application [31].

Artificially Expanded Genetic Information System (AEGIS) represents a groundbreaking advancement in synthetic biology, designed to transcend the limitations of these early synthetic systems (Fig. 1a). AEGIS introduces additional synthetic nucleotides by rearranging hydrogen bond donor and acceptor groups while retaining their Watson-Crick geometries [35]. These artificially created bases can form complementary pairs, just like natural bases, but they provide new pairing options that increase the coding capacity of genetic sequences [36]. AEGIS enhances the capabilities of XNA by increasing information density and functional diversity [31]. For instance, this expanded genetic alphabet improves the predictability of XNA self-assembly and enables the creation of more effective aptamers through in vitro selection processes, which bind specifically to breast cancer cells [36]. The OligArch software facilitates the design of self-assembling XNA constructs using six- and eight-letter AEGIS [37]. This expansion allows for the creation of novel proteins and biological functions that are not possible within the constraints of natural genetic codes [38]. AEGIS has significant implications for biotechnology and medicine, offering potential applications in areas such as drug delivery and cancer therapy [39]. By augmenting the genetic toolkit available to scientists, AEGIS opens up unprecedented opportunities for innovation in synthetic biology [39,40].

While AEGIS offers potential for more complex information systems, the field is still in its early stages requiring collaborative efforts from multiple disciplines, such as synthetic biology, biotechnology, and computational biology, to fully realize the use of expanded genetic codes (Fig. 1b). Structural and computational studies have provided insights into the conformational dynamics of AEGIS XNA, revealing differences from Watson-Crick DNA and paving the way for developing technologies that can manipulate AEGIS nucleobase pairs [41]. The ability to encode and manipulate non-standard nucleotides allows for the development of novel therapeutic interventions and diagnostic tools such as new aptamer technologies [31,36]. Additionally, in synthetic biology, expanded genetic systems can be harnessed to create synthetic organisms or biomolecules with tailored functions such as semi-synthetic *Escherichia coli* [38,42,43]. By incorporating additional base pairs into the genetic code, researchers can encode more complex biological functions, opening up possibilities for biotechnological and medical applications.

## Nanopore Sequencing's Unique Potential for Non-Standard Nucleotides

Artificially Expanded Genetic Information Systems (AEGIS) offer enhanced functionality and information density compared to natural DNA, but sequencing these non-standard nucleotides presents challenges [44]. Traditional sequencing technologies cannot sequence non-standard nucleotides directly due to the reliance on amplification steps that require modified polymerase enzymes and specific non-standard AEGIS components [39]. Thus, controlled transliteration of AEGIS DNA to standard DNA has been at the core of sequencing non-standard nucleotides, and recent advancements have improved sequencing capabilities for AEGIS. For instance, the Enzyme-Assisted Sequencing of Expanded Genetic Alphabet (ESEGA) method enables high-throughput sequencing of 6-letter AEGIS XNA at single-base resolution [39]. This approach facilitates the optimization of 6-nucleotide PCR conditions and the evaluation of DNA polymerase fidelity with AEGIS components. Earlier work demonstrated the amplification and sequencing of 6-letter AEGIS XNA with low mutation rates, allowing the artificial genetic system to evolve without reverting to natural DNA [45].

Nanopore sequencing bypasses these limitations, allowing for the direct and continuous readout of long XNA strands. Currently, the range of read length varies from the shortest read length of 20 b to the ultra-long read length of 3 megabases (Mb) DNA [46] and 20 kilobases (kb) RNA [47]. Furthermore, nanopore sequencing does not require amplification, enzymatic synthesis, or nucleotide modification [48–50]. A nanopore detects changes in ionic current as a biomolecule translocates through it, with the magnitude of the current and the duration of the translocation depending on the geometry, size, and chemical composition of the molecule [12,51]. Since these biophysical properties vary across different non-standard nucleotides (Fig. 1a), nanopore sequencing is particularly advantageous for synthetic genomes or expanded genetic systems, where conventional technologies often struggle to identify or interpret novel base pairings [48]. A recent study of applying nanopore sequencing to the expanded genetic code demonstrated a broader signal range than standard DNA alone and single-base substitutions with AEGIS are distinguishable with high confidence [48]. However, challenges remain, such as reduced processivity of helicases with C-glycoside bases from AEGIS [44]. These developments in XNA sequencing support the continued exploration of expanded genetic codes in biotechnology applications.

The flexibility of output signals and downstream analysis from nanopore sequencing further enhances its suitability for non-standard nucleotides, allowing for real-time interpretation of diverse sequences. However, decoding output signals from nanopore sequencing scales exponentially with the number of nucleotide types in a genetic code. For example, decoding a 6-base $k$-mer model for a 4-alphabet genetic code requires processing 4,096 unique $k$-mers, whereas decoding the same 6-base $k$-mer model for an 8-alphabet genetic code requires processing 262,144 unique $k$-mers [48]. To overcome this hurdle in XNA data processing, computational complexity may be reduced by exploiting the capacity of nanopore technology to selectively sequence specific regions of interest (Fig.

1b). Furthermore, applying machine learning algorithms for sequence decoding is poised to handle complex datasets generated by expanded genetic codes. The ability to sequence without prior knowledge of the nucleotide composition also allows nanopore sequencing to function effectively in exploratory research where both standard nucleotides and non-standard nucleotides may be present.

## CONCLUSION

Looking ahead, the integration of nanopore sequencing with other emerging technologies such as artificial intelligence and synthetic biology promises to accelerate our understanding of non-standard genetic systems. As nanopore technology continues to evolve, its applications will likely extend beyond natural genetic systems, becoming a cornerstone in biotechnology, medicine, and even astrobiology. To fully unlock the potential of nanopore sequencing for non-standard nucleotides, several key advancements are necessary. Improved data processing algorithms, specifically designed to recognize a broader range of biochemical structures, will be essential. Additionally, developing reference databases for non-standard nucleotides using short-read sequencing or long-read sequencing will facilitate more accurate biological interpretations aforementioned in this perspective. Nanopore sequencing has already transformed our ability to sequence long stretches of natural genetic sequences, but its futuristic potential lies in the ability to decode non-standard nucleotides. As we expand the genetic code and discover its relevance in multiple disciplines, nanopore technology stands at the forefront, offering a robust platform for deciphering complex, novel biological information. By harnessing its unique capabilities, we can unlock new frontiers in synthetic biology, bioengineering, and beyond.

## ACKNOWLEDGMENTS

The research and development activities described in this study were funded by CSU, Fresno. The author would like to thank the members of the Department of Biology at CSU, Fresno and the organizers of the Jagna International Workshop.